
\input harvmac.tex
\overfullrule=0pt

\def\GeV{\rm GeV}

\Title{\vbox{\baselineskip12pt\hbox{BUHEP-95-18}}}
{\vbox{\centerline{Technipion contribution to $ b\rightarrow s\gamma$}}}

\centerline{Bhashyam Balaji}
\bigskip\centerline{Physics Department}
\centerline{Boston University}\centerline{Boston, MA 02215}
\vskip .3in

\centerline{\bf Abstract}
\bigskip
We show that the present limit on the inclusive decay $b\rightarrow s\gamma$
provides strong constraints on Technicolor models. In particular, small
values of $F_\pi$ and the mass of charged octet and singlet technipions
are excluded, assuming the most natural form of the technipion
coupling to the ordinary quarks.

\Date{5/95}

\vfil\eject

The $b\rightarrow s\gamma$ decay is very sensitive to the physics beyond
the standard model
\ref\hew{ J. L. Hewett,  SLAC-PUB-6521}
. The one-loop W-exchange diagrams that generate this decay at the
lowest order in the standard model are shown in Figure 1. In most extensions of
the standard model (such as the two-higgs-doublet models), there are
additional contributions to this process coming from charged scalar
exchange (replacing the W with the charged scalar(s) in Figure 1).
This has been used to obtain lower bounds on the masses of the charged scalars.

Technicolor theories
\ref\SWLS{ S. Weinberg,  Phys. Rev. {\bf D19} (1979) 1277;
L. Susskind,  Phys. Rev. {\bf D20} (1979) 2619}
are an attempt to explain electroweak symmetry breaking without
elementary scalars. Non-minimal models
can contain pseudo-Goldstone bosons (technipions),
which arise from the breakdown of global chiral
symmetries
\ref\EL{ E. Eichten and K. D. Lane, Phys. Lett. {\bf 90B} (1980) 125 }
. Here we consider the  contribution of
color singlet and color octet, weak isotriplet
technipions to the process
$B\rightarrow X_s\gamma$.
%
%

Randall and  Sundrum have studied the short-distance  extended
technicolor
(ETC) \EL,
\ref\DS{ S. Dimopoulos and L. Susskind, Nucl. Phys. {\bf B155} (1979)
237}
 contribution to $b\rightarrow s\gamma$,
that comes from the single ETC gauge boson exchange connecting purely
left-handed doublets ( Figure 2)
\ref\LRRS{ L. Randall and R. Sundrum, Phys. Lett. {\bf B312} (1993) 148 }
. They show that this contribution is
suppressed relative to the standard model (SM) one by
$m_t / 4\pi v $, where $v = 246 {\rm GeV}$. We shall see that the
leading contributions are long-distance, nonanalytic contributions from
technipion exchange, and that they are comparable to that from W-exchange.
These contribution will be used
obtain bounds on masses of the charged color octet and color singlet
technipions.

Consider an ETC model incorporating the one
family technicolor model.
The exchange of ETC gauge bosons induces interactions of the form
$$\eqalign { {g_{ETC}^2\over M_{ETC}^2} Y^{ij}_u( \overline{\psi} _L^i \gamma
_{\mu} T_L )(\overline U _R \gamma ^{\mu} u^j _R ) +
{g_{ETC}^2\over M_{ETC}^2} Y^{ij}_d ( \overline{\psi} _L^i \gamma _{\mu} T_L )
(\overline D _R \gamma ^{\mu} d^j_R) +h.c. }\eqno(1)$$
where $g_{ETC}$ is the ETC gauge boson coupling and $M_{ETC}$ the generic
gauge boson
mass. Here, $T_L$ is the colored techni-doublet while $U_R$, $D_R$ are
the right-handed partners; $\psi ^i_L$ is the $i$-th generation
quark doublet and $u^i_R$, $d^i_R$ are its right-handed partners.
The
matrices $Y_{u(d)}$ are model dependent constants that
play a role similar to the standard model Yukawa couplings.

The relevant  effective  ETC  interactions of the technipions
and the top and bottom quarks (ignoring terms proportional to
the CKM matrix element $V_{td}$)
are obtained from PCAC arguments using the above Lagrangian.
They have  the form, respectively,
$$\eqalign { c_1 {\pi ^+ _{singlet}  \over {\sqrt6 F_{\pi}}}( m_t (
\overline t _R V_{tb} b_L +
\overline t _R V_{ts} s_L ) - m_b ( \overline t _L V_{tb} b_R +
\overline t _L V_{ts} s_R ) + h.c. + \cdots )}\eqno(2a)$$
and
$$\eqalign { c_8 {\pi ^+ _{octet} \over F_{\pi}} ( m_t ( \overline t _R
V_{tb} {\lambda ^a \over 2} b_L +
\overline t _R V_{ts} {\lambda ^a \over 2} s_L )
- m_b ( \overline t _L V_{tb} {\lambda ^a \over 2} b_R +
\overline t _L V_{ts} {\lambda ^a \over 2} s_R ) + h.c. + \cdots )}\eqno(2b)$$
for the color singlet and color octet technipions --- the only
technipions that couple to ordinary (color-triplet) quarks
\ref\Ell{ J. Ellis  {\it et al.}, Nucl. Phys. {\bf B182} (1981) 529 }
. The $\lambda ^a$'s
are the SU(3) generators normalised so that  $ tr(\lambda ^a \lambda ^b)
= 2\delta ^{ab}$. The constants $c_1$, $c_8$ are model dependent factors
of of order 1. We shall take $c_{1,8} = 1$;
alternatively, all our results are really functions of $F_\pi /c_{1,8}$,
rather than $F_\pi$.

The couplings of the charged technipions to the quarks
are  similar to that of the  charged Higgs coupling to the quarks
in two Higgs doublet model of type I in which
both the up- and down-type
quarks get mass from Yukawa couplings to the same Higgs doublet
\ref\mod1{ H. E. Haber, G. L. Kane, and T. Sterling, Nucl. Phys.
 {\bf B161} (1979) 493; L. Hall and M. Wise, Nucl. Phys. {\bf B187}
(1981) 397 }
. In particular, it is important
to observe the relative negative sign between the terms proportional to
$m_t$ and on $m_b$.
This  sign is model-dependent. For  instance, the charged technipions
in a two-doublet technicolor model (with $SU(4)\times SU(4)$ chiral symmetry)
couple to ordinary quarks  without the relative sign difference if
one of the doublets is used for giving mass to the up-type quarks and
the other to the down type quarks (if the same doublet is used, one
reproduces Eq. 3). We shall only study the case when the relative sign
is negative as that is most conservative for the lower bounds
on technipion masses.

We are interested in the technipion contribution to
the coefficient $C_7$ of the
operator known as $\hat{O}_7$ in the notation of
\ref\GrSpWi{ B. Grinstein, R. Springer, M. B. Wise,  Nucl. Phys.
{\bf B339} (1990) 269 }
$$\eqalign { \hat{O}_7 = {\left( 4G_F\over {\sqrt2} \right)} V^*_{ts} V_{tb}
 {\left({e\over {16\pi ^2}}\right) m_b({\overline s_L} \sigma ^{\mu \nu}
b_R ) F_{\mu \nu}}}\eqno(3)$$
where $F_{\mu \nu}$ is the electromagnetic field strength.
The relevant Feynman diagrams are obtained by replacing the W with the
charged technipions in Figure 1.
The virtual top quark contribution is the most
important one since the technipion coupling to quarks is expected to be
proportional to the quark mass.

Comparing with the results of
\ref\GrWi{ B. Grinstein and M. B. Wise,  Phys. Lett. {\bf B201} (1988)
274 }
\ref\HoWi{ W. S. Hou and R. Willey,  Phys. Lett. {\bf B202} (1988)
591 }
for the two Higgs-doublet models
, the technipion contributions to $C_7$ are found to be
$$\eqalign{C_7 ({\pi _{singlet}}) = + {\sqrt2 \over {4 G_F F_\pi
^2}}{\left({1\over 6}\right)}
\left(B(x)-
{A(x)\over 6}\right)}\eqno(4a)$$
and
$$\eqalign{C_7 ({\pi _{octet}}) = + {\sqrt2 \over {4 G_F F_\pi ^2}}
{\left({4\over 3}\right)}
\left(B(x)-
{A(x)\over 6}\right)}\eqno(4b)$$
where $x = (m^2_t / M_{\pi _T} ^2)$ and
$$\eqalign{A(x) = x\left({{{2\over 3} x^2 + {5\over 12} x - {7\over 12}}\over
{(x-1)^3}} -
{{({3\over 2} x^2 - x) \ln x}\over {(x-1)^4}}\right)}\eqno(5)$$
$$\eqalign{B(x) = {x\over 2}\left({{{5\over 6} x -{1\over 2}}\over {(x-1)^2}} -
{{(x-{2\over 3})\ln x}\over {(x-1)^3}}\right)}\eqno(6).$$
For comparison, the W-exchange
contribution to $C_7$
that is present in SM as well as technicolor models
is $ (- A(x)/2)$
\ref\InLi{ T. Inami and C. S. Lim,  Progr. Theor. Phys.  {\bf 65}
(1981) 297
}
if QCD corrections are ignored
. The sign of the technipion
contribution is positive and twnds to decrease the overall magnitude
of $C_7$ when added to  the SM
contribution.

There are many theoretical and experimental uncertainities  present in the
prediction for BR[B$\rightarrow X_s\gamma$]
\ref\Buras{ A. J. Buras, M. Misiak, M. M\"unz and S. Pokorski,
Nucl. Phys.  {\bf B424 } (1994) 374}
. The two most important sources of uncertainities are
the uncertainity in $\alpha _s$
\ref\LaPo{ P.Langacker and N. Polonsky,  Phys. Rev. {\bf D47}
 (1993) 4028 }
and next-to-leading-order QCD effects \Buras .
We incorporate the uncertainity in $\alpha _s (M_z) $ by
obtaining results for $\alpha _s (M_z) = 0.11$ and
$\alpha _s (M_z) = 0.13 $.
By varying the renormalization scale $\mu$ by a factor of $2$ in both
directions around $5$ $\GeV$ we estimate the size of the next-to-leading
QCD corrections --- a $25\%$ effect on $C_7$\Buras .

We  ignore uncertainities in ${m_c/m_b}$
\ref\Ruckl{ R. R\"uckl,  MPI-Ph/36/89}
, the top quark mass
\ref\top{ F. Abe {\it et al}. (CDF Collaboration), Phys. Rev. {\bf D50}
(1994) 2966; Phys. Rev. Lett. {\bf 73} (1994) 225; S. Abachi {\it et
al.} (D0 Collaboration), FERMILAB-PUB-95-028-E
}
, the ratio of CKM factors ${{\mid V^*_{ts} V_{tb} \mid  ^2}/
{\mid V_{cb} \mid ^2}}$, the experimental determination  of
BR[B $\rightarrow X_e e \overline{\nu _e}$]
\ref\PDB{ Review of Particle Properties, Phys. Rev. {\bf D45} 1994}
and the spectator model
approximation. These effects are expected to give about a $15\% $ change in
our calculation of the result. This has a negligible
impact on the lower bounds on $M_{\pi _T}$ (as a function of $F_{\pi}$),
as the technipion contribution to $C_7$ is opposite in sign to the
SM contribution.
We use the central values of  these other parameters  :
$m_t = 175$ $\GeV$ \top , ${z= {m_c/m_b}} = 0.316$
\Ruckl \    and BR[B$\rightarrow X_c e\overline{\nu _e}$] $= 10.7\%$ \PDB .
We take the central value of $0.95$ for ${{\mid V^*_{ts} V_{tb} \mid ^2}/
{\mid V_{cb} \mid ^2}}$ in our computation of the
standard model's  leading logarithmic contribution \GrWi ,
\ref\SML{  R. Grigjanis, P. J. O' Donnell, M. Sutherland and H. Navelet,
Phys. Lett. {\bf B213} (1988) 355,  Phys. Lett. {\bf B286}
(1992) 413; G. Cella, G. Curci, G. Ricciardi and A. Vicere,
 Phys. Lett.  {\bf B248} (1990) 181; M. Misiak,  Phys. Lett.
{\bf B269} (1991) 161; M. Misiak,  Nucl. Phys.  {\bf B393} 23;
K. Adel and Y. P. Yao,  Modern Physics Letters {\bf A8} (1993)
1679; M. Ciuchini {\it et al }, Phys. Lett. {\bf 316} (1993)
127}
. The limits on the branching ratio ( each at 95\% CL )
of  $B\rightarrow X_s \gamma $ are  $1.0\times 10^{-4} <$
BR[$B\rightarrow X_s \gamma $] $<
4.2\times 10^{-4}$
\ref\Alam{ M.S. Alam {\it et al} ( CLEO Collaboration ),
Phys. Rev. Lett. {\bf 74} 2885 (1995)}
, which correspond to $0.18\leq \vert C_7 \vert \leq 0.38$.

The results obtained for the color octet and color singlet cases
are shown in Figs 3 and 4.  The standard model contribution
from W exchange is added to the technipion contributions and the
sum is compared with the experimental limits
in order to obtain the bounds on the masses of the technipions.
The  lower bound is
larger in the case of the color octet because of the larger  group theory
factor. The   next-to-leading
order contributions do not affect the lower bounds on technipion masses
significantly
(again because the technipion contribution to $C_7$ interferes destructively
with the SM contribution). However, they do affect the
excluded region that comes  from the  experimental lower bound on
BR[B$\rightarrow X_s \gamma $] (the shaded region in between).
This is because the technipion contribution is positive definite and
there is a lower limit on $\vert C_7 \vert $. Consequently, there
is a certain region which is disallowed  as $\vert C_7 \vert$
(which includes the W and technipion contributions) would then be
smaller than the expermental lower bound $0.18$. This region is
thus sensitive to the next-to-leading order contributions.
We have shown the results from the two
extreme cases that  could result from the uncertainity in $\alpha _s (M_Z)$
and  the choice of appropriate $\mu $.

A one-family technicolor model has contributions from
both the color octet and color singlet technipions.
Both contributions are positive definite and hence only serve to
increase the lower bounds on technipion mass for a given $F_\pi$.
In Fig. 5 we have plotted the excluded(shaded) regions in the
$M_{\pi _{octet}} - M_{\pi _{singlet}}$ plane in a one-family
technicolor model with $F_\pi =125 $ and $c=1$.
The unshaded region in between is the allowed region of
physical interest. For instance, for a color singlet technipion of mass
of about 100 GeV, the octet mass has to be between 200 and 350 GeV.
As above, the shaded region  in between
is due to the experimental lower bound on $\vert C_7(\mu) \vert $.

Our conclusions depend on the relative negative sign in Eq. (2);
$C(\pi _T)$ would be negative definite if there were no relative
sign difference. This is because the $B(x)$ piece \ \ \
(which is the
contribution from the `cross term' in Eq. (2) ) dominates.
In a more general technicolor model with more technipions
, it is apparent that
the lower limits on the technipion masses would be higher than here
if all the relevant technipions couple as in
Eq. (2), i.e., with a  relative negative sign.  If all of them coupled with a
relative positive sign, the situation is even worse, since then the
contribution to $C_7 $ would be negative, like the standard model result.
However, the constraint would
be relaxed (relative to the above) if  some of the  technipions
coupled to quarks with a
relative negative sign
and the others with a relative positive sign, so that
there were some cancellation.

In the Topcolor assisted Technicolor model proposed by Hill
\ref\chill{ C. T. Hill, Phys. Lett. {\bf B345}(1995) 483}
, top-pions ( $\tilde{\pi} $ ) can contribute to $b\rightarrow s\gamma$.
The relevant term in the interaction lagrangian is
$$\eqalign{ { m_t\over f_{\pi}} \tilde{\pi} ^+ \left[ ( \overline t _R s_L
D_{Lbs}  + \overline c _R b_L U^*_{Rtc} ) + h.c. \right]}\eqno(7)$$
where $U_{L,R}$ and $D_{L,R}$ are
the field redefinition matrices for the up-type (down-type)
quarks. The top-pions couple only to the third generation weak doublet.
By assigning different quantum numbers to the first two generations
and the third generation, the third generation is distinguished
from the other two. Then the $U_{L,R}$ and  $D_{L,R}$ cannot be
completely absorbed into the CKM matrix, as in the standard model
or the ETC interaction assumed in Eq. 1.  As the elements of $D_L$  is not
measurable (only the CKM matrix $V = U^\dagger _L D_L $ is),
no definite bounds on the top-pion mass can be obtained (see \chill ).

An ETC model in which the third generation is distinguished from the
other two would also involve  experimentally undetermined
parameters (in place of the CKM matrix elements), and hence be less
severely constrained than the type of models studied here.
In the case when ETC is used in conjunction with topcolor, the bounds
would be considerably weaker for the technipions,
because the technipion coupling is proportional to the ETC
generated  part of the top quark mass.
In topcolor assisted technicolor models, the ETC
contribution to the top quark mass $\epsilon m_t$ is small,
i.e., $\epsilon \ll 1$.

\bigskip
\medskip

The author wishes to thank R. S. Chivukula and K. Lane for suggesting the
investigation and guiding it.
He also thanks C. Hill and D. Kominis for useful
discussions on the Topcolor model and M. V. Ramana, E. H. Simmons
and J. Terning for comments on
the manuscript. The work was supported by the Department of Energy under
Grant No. DE-FG02-91ER40676.

\listrefs

\vfil\eject

\centerline{\bf Figure Captions }
\bigskip

\item{[1]} Feynman diagrams that determine the one-loop $b\rightarrow
s\gamma $ decay amplitude. The loop with virtual top quark dominates.
\item{[2]} The ETC gauge boson exchange contribution to the $b\rightarrow
s\gamma $ considered by Randall and Sundrum.
\item{[3]} The excluded regions (shaded) in the $F_\pi $-$M_\pi $ plane for the
color singlet technipion. Figures 1(a) and 1(b) correspond to the
$[ \alpha _s (M_z), \mu $ in GeV ] values of $[ 0.13, 2.5 ]$ and
$[ 0.11, 10 ]$ respectively.
\item{[4]} The excluded regions (shaded) in the $F_\pi $-$M_\pi $ plane
for the color octet technipion. Figures 2(a) and 2(b) correspond to the
$[ \alpha _s (M_z), \mu $ in GeV ] values of $[0.13, 2.5 ]$ and
$[ 0.11, 10 ]$ respectively.
\item{[5]} The excluded  regions (shaded)
in the $M_{\pi _{octet}} - M_{\pi _{singlet}}$
plane for $F_\pi $ = 125 GeV . We take $\alpha _s (M_z) =0.11$ and
$\mu = 10$ GeV.

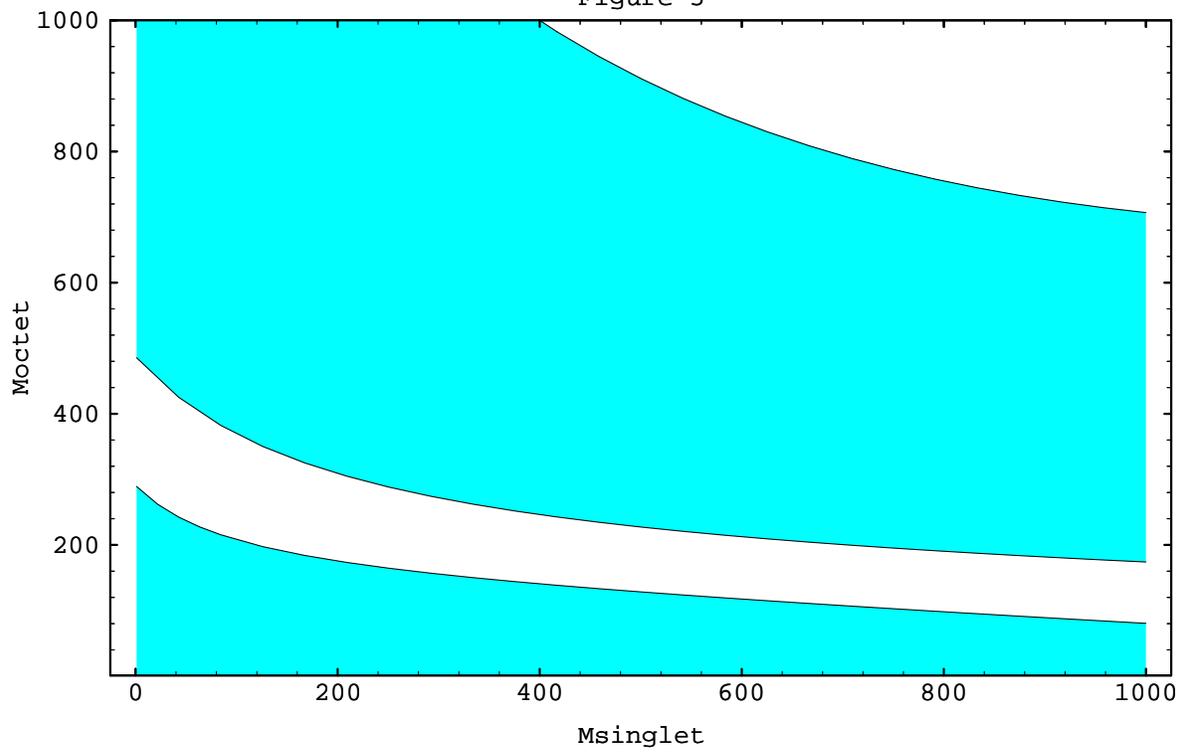
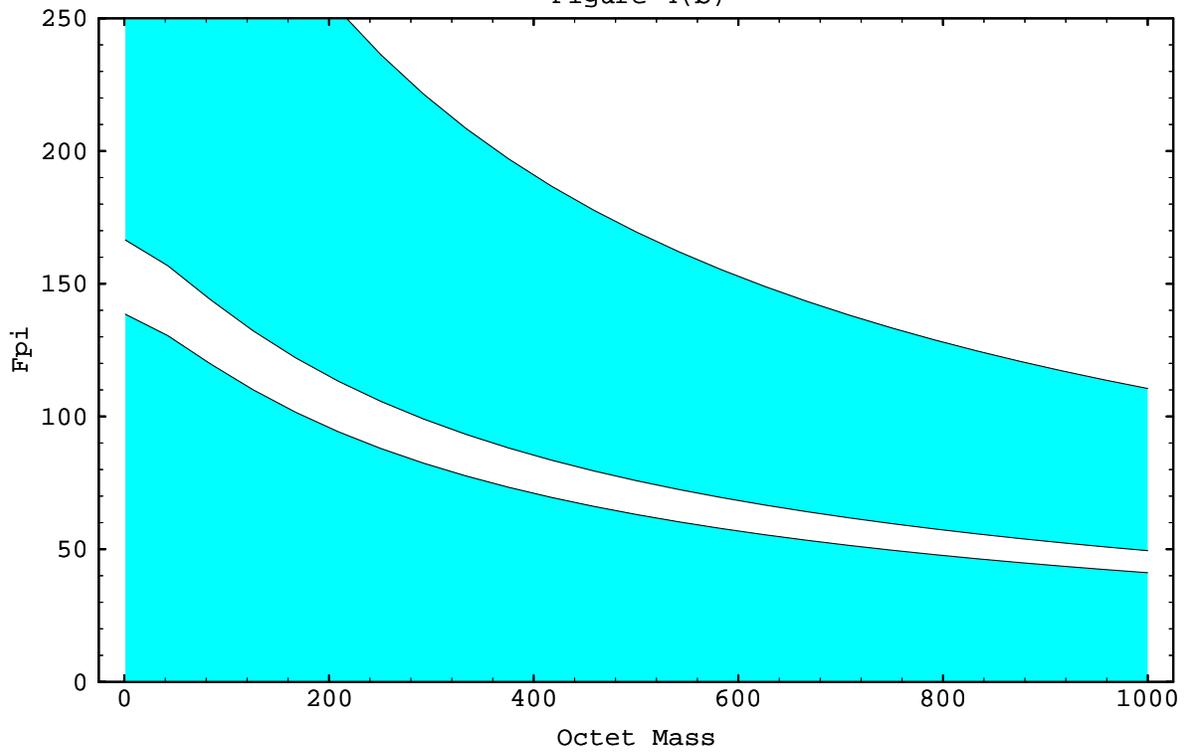
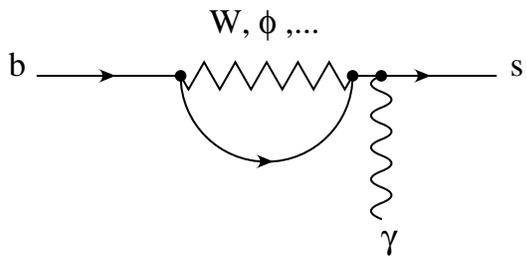
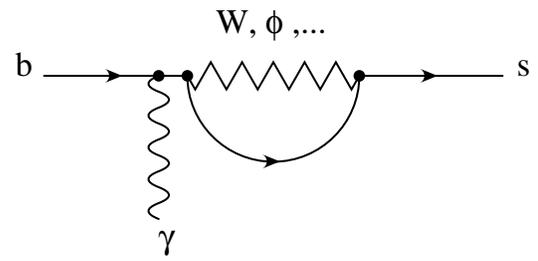
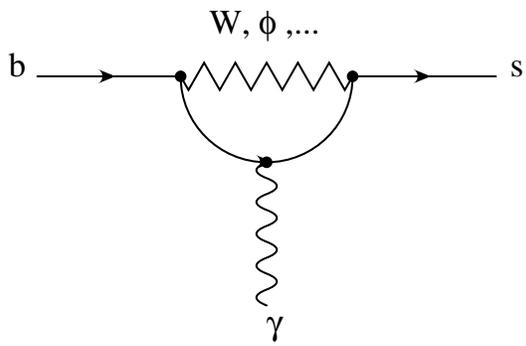
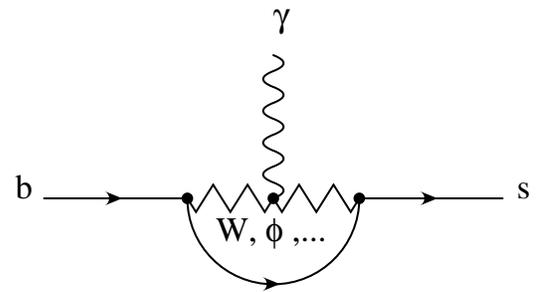
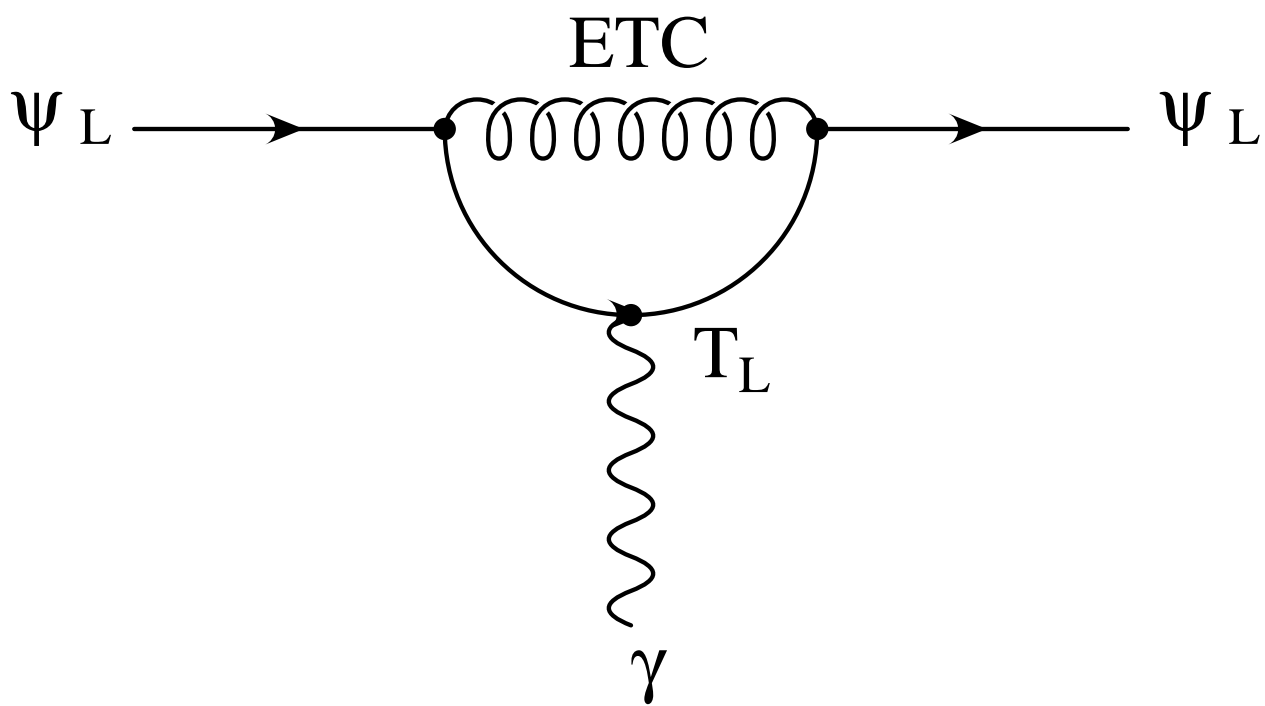
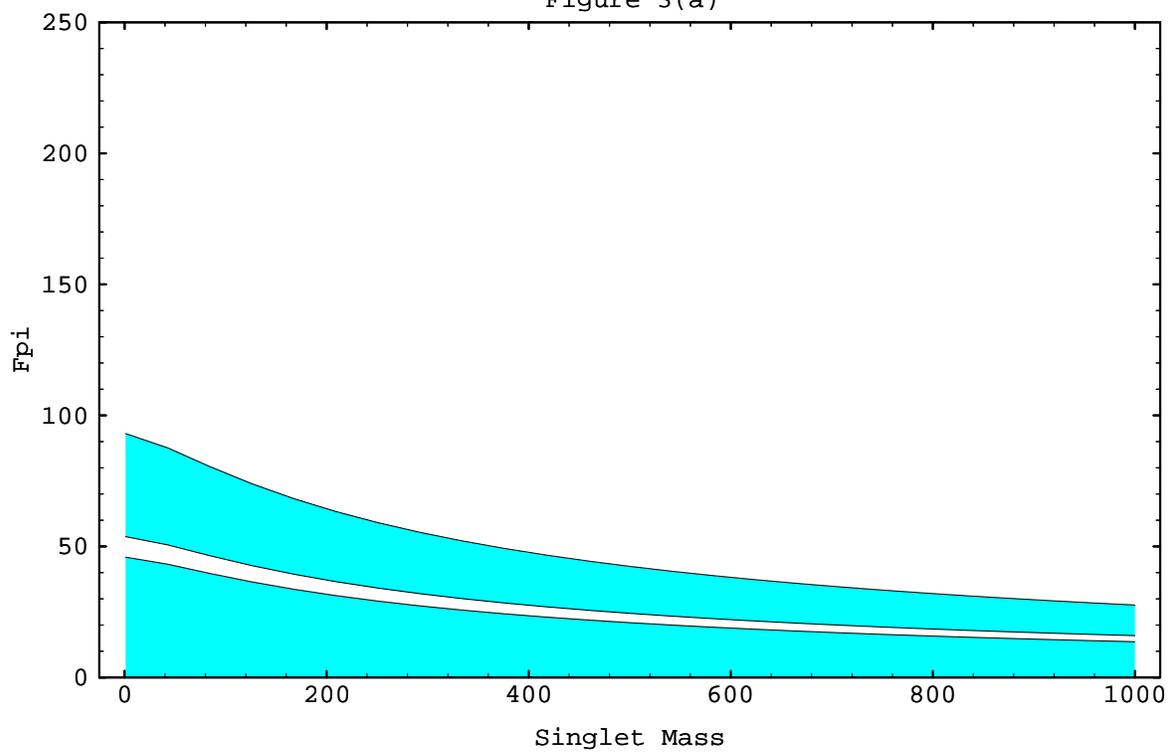
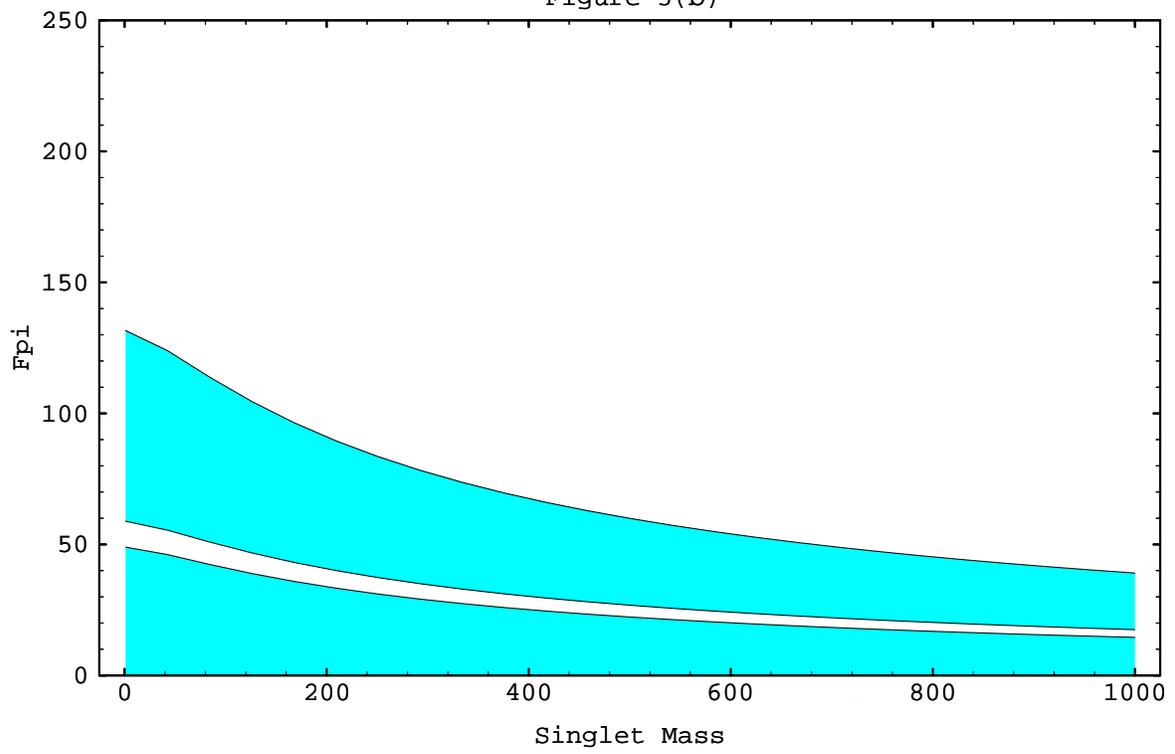
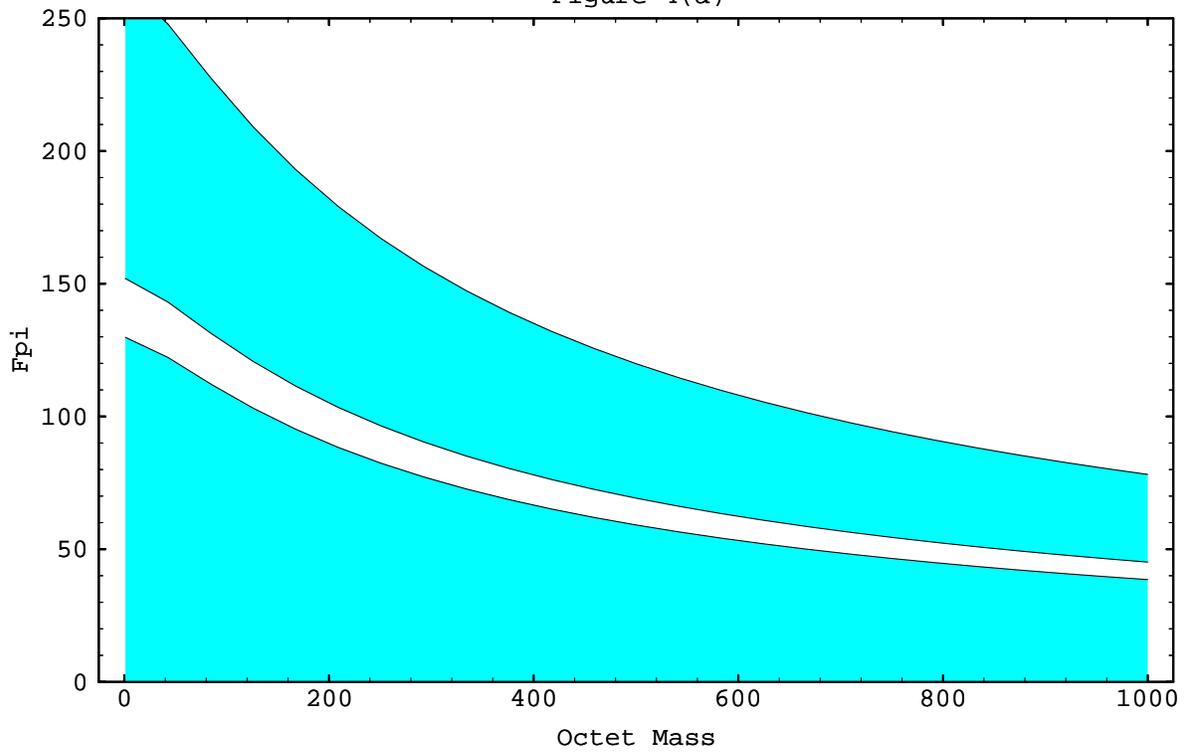
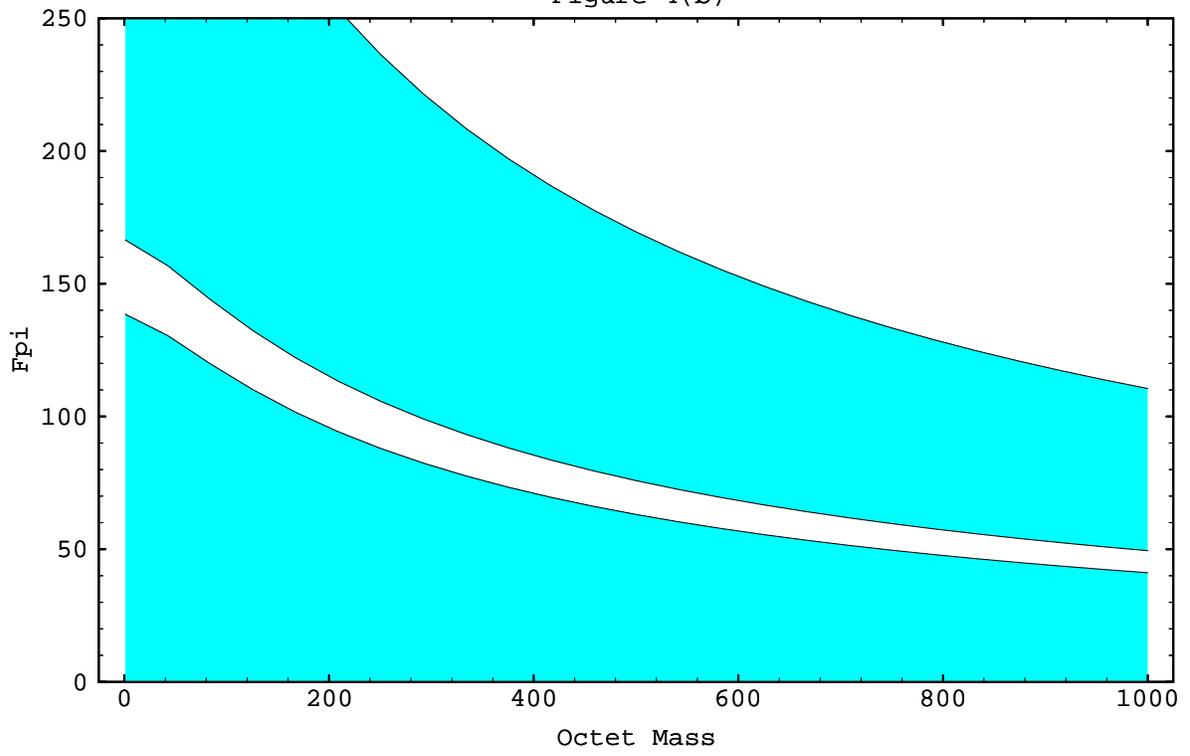
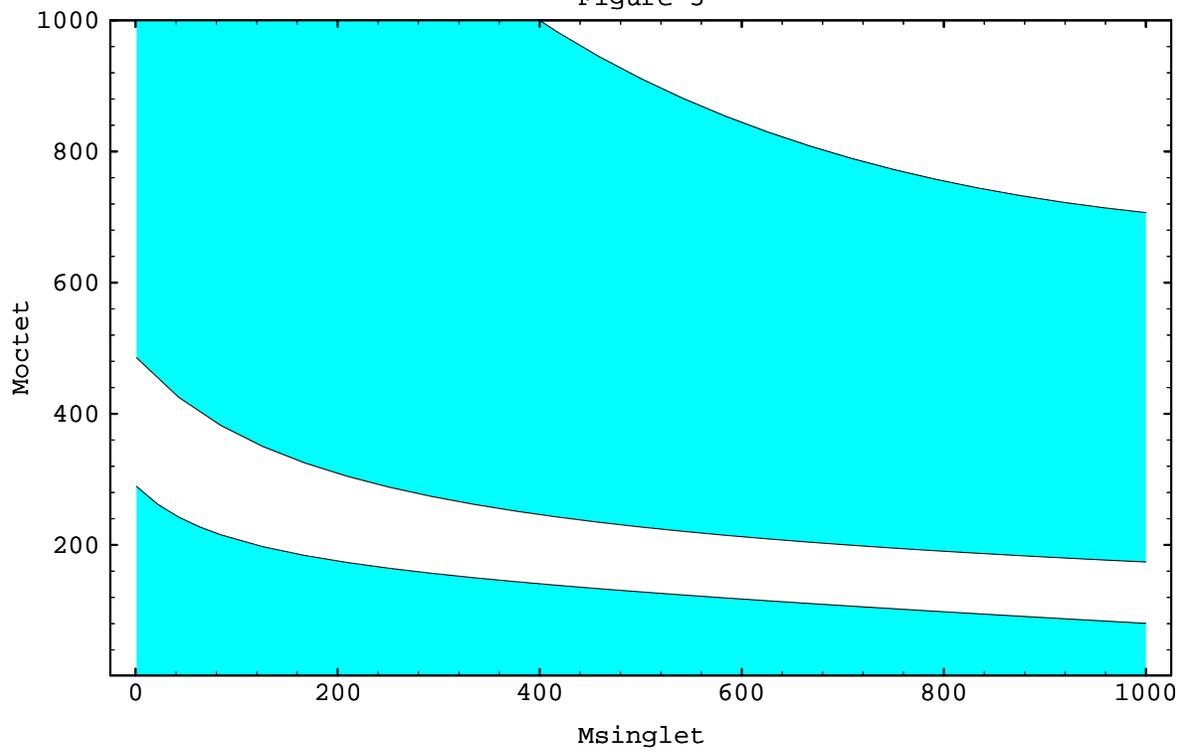
\bye